\documentclass[a4paper,11pt]{elsarticle}

\makeatletter
\def\ps@pprintTitle{%
 \let\@oddhead\@empty
 \let\@evenhead\@empty
 \def\@oddfoot{\centerline{\thepage}}%
 \let\@evenfoot\@oddfoot}
\makeatother

\usepackage{amssymb,amsthm,mathtools}
\usepackage{resizegather,xcolor}
\usepackage[left=20mm,right=20mm,bottom=25mm]{geometry}

\usepackage[colorlinks]{hyperref}
\AtBeginDocument{%
  \hypersetup{citecolor=black!50!green,
              linkcolor=black!10!red}}

\usepackage{enumitem}

\newtheorem{lem}{Lemma}
\newtheorem{cor}{Corollary}
\newtheorem{prop}[cor]{Proposition}

{\theoremstyle{definition}
\newtheorem{rmk}{Remark}}

{\theoremstyle{definition}
\newtheorem{example}{Example}}

\newdefinition{defi}{Definition}
\newproof{prf}{Proof}

\newcommand{\GF}[1]{\mathbb{F}_{2^{#1}}}
\newcommand{\Tr}{\mathrm{Tr}}
\newcommand*{\QED}{\hfill\ensuremath{\square}}

\newcommand{\order}[1]{\left|#1\right|}

\begin{document}

\allowdisplaybreaks[4]

\begin{frontmatter}

\title{On the Derivative Imbalance and Ambiguity of Functions}

\author[amss]{Shihui Fu}
\ead{fushihui@amss.ac.cn}
\author[amss,stcsl]{Xiutao Feng}
\ead{fengxt@amss.ac.cn}
\author[carleton]{Qiang Wang}
\ead{wang@math.carleton.ca}
\author[laga]{Claude Carlet}
\ead{claude.carlet@univ-paris8.fr}

\address[amss]{KLMM, Academy of Mathematics and Systems Science, Chinese Academy of Sciences, Beijing 100190, CHINA}
\address[stcsl]{Science and Technology on Communication Security Laboratory, Chengdu 610041, CHINA}
\address[carleton]{School of Mathematics and Statistics, Carleton University, Ottawa ON K1S 5B6, CANADA}
\address[laga]{LAGA, University of Paris 8, Saint-Denis Cedex 02, FRANCE and University of Bergen, NORWAY}

\begin{abstract}
In 2007, Carlet and Ding introduced two parameters, denoted by $Nb_F$ and $NB_F$, quantifying respectively the balancedness of general functions $F$ between finite Abelian groups and the (global) balancedness of their derivatives $D_a F(x)=F(x+a)-F(x)$, $a\in G\setminus\{0\}$ (providing an indicator of the nonlinearity of the functions). These authors studied the properties and cryptographic significance of these two measures. They provided for S-boxes inequalities relating the nonlinearity $\mathcal{NL}(F)$ to $NB_F$, and obtained in particular an upper bound on the nonlinearity which unifies Sidelnikov-Chabaud-Vaudenay's bound and the covering radius bound. At the Workshop WCC 2009 and in its postproceedings in 2011, a further study of these parameters was made; in particular, the first parameter was applied to the functions $F+L$ where $L$ is affine, providing more nonlinearity parameters.

In 2010, motivated by the study of Costas arrays, two parameters called ambiguity and deficiency were introduced by Panario \emph{et al.} for permutations over finite Abelian groups to measure the injectivity and surjectivity of the derivatives respectively. These authors also studied some fundamental properties and cryptographic significance of these two measures. Further studies followed without that the second pair of parameters be compared to the first one.

In the present paper, we observe that ambiguity is the same parameter as $NB_F$, up to additive and multiplicative constants (i.e. up to rescaling). We make the necessary work of comparison and unification of the results on $NB_F$, respectively on ambiguity, which have been obtained in the five papers devoted to these parameters. We generalize some known results to any Abelian groups and we more importantly derive many new results on these parameters.

\end{abstract}


\end{frontmatter}

\section{Introduction}\label{secn:Introduction}

Functions between two finite Abelian groups play a very important role due to their applications in combinatorics, error correcting coding theory, and cryptography. In combinatorial design, the graph of each perfect nonlinear function from a finite Abelian group to one of its normal subgroups can give rise to a semiregular relative difference set \cite{DAM:Pott04}. In error correcting coding theory, every binary linear code of dimension $m$ and length $2^n$ for some positive integers $n$ and $m$, can be associated with an $(n,m)$-function (the code being identified with the set of its component functions), and vice versa. More generally, important linear or unrestricted codes (Reed-Muller, Kerdock codes \cite{Codes:MacWilliamsS77}) are defined as sets of Boolean functions. In modern cryptography, confusion and diffusion are two fundamental properties of secure ciphers identified by Shannon. Confusion is reflected in the nonlinearity (with diverse meanings of this term, the parameter explicitly called nonlinearity being more related to the linear attack - see below - and other nonlinearity parameters being related to other attacks) of the primitives in the cryptosystem which are not linear, since linear systems are generally easy to break. Currently, since vectorial Boolean functions can easily provide confusion, they are commonly used to serve as cryptographic primitives, for instance, as Substitution boxes (S-boxes), to make a system secure. AES is an example, which uses a function from $\GF{}^8$ to $\GF{}^8$, parallelized 16 times and composed with different linear permutations, to serve as its nonlinear part. Another well-known example is the Data Encryption Standard (DES), which uses eight S-boxes and each is a map from $\GF{6}$ to $\GF{4}$. While most modern cryptosystems use S-boxes that are based on vectorial Boolean functions, there are situations (encrypting credit card numbers or social security numbers, for example) where nonbinary data is a natural part of the application and one might use nonbinary functions in the cryptosystem. For example, the Exponential Welch Costas (EWC) functions from $\mathbb{Z}_{256}$ to itself, as well as their inverses, the Logarithmic Welch Costas (LWC) functions are used as S-boxes in SAFER family of cryptosystems, proposed by Massey \cite{FSE:Massey93,FSE:Massey94}. All of these functions can be viewed as maps between two finite Abelian groups with possible different orders. This is one of the motivations for studying the maps between any two finite Abelian groups.

The differential attack, introduced by Biham and Shamir \cite{JOC:BihamS91}, successfully applies when any two plaintexts with fixed difference lead after the last-but-one round to outputs whose difference takes a certain value with a probability significantly larger than the uniform probability. The larger the probability of the differential, the more efficient is the attack. The related criterion on a function $F$ from $G_1$ to $G_2$ used as an S-box in the round functions of the cipher is that the output of its derivative
\[
D_a F(x)=F(x+a)-F(x)
\]
at any nonzero $a\in G_1$ must be as uniformly distributed as possible.

Another  most prominent attack is the linear cryptanalysis introduced by Matsui \cite{EUROCRYPT:Matsui93}. The nonlinearity $\mathcal{NL}(F)$ of a function $F$ quantifies its resistance to this kind of attack. This parameter is equal to the minimum distance from the function to all affine functions. In the case of functions from $\GF{}^n$ to $\GF{}^m$, those attaining the maximum nonlinearity are called {\em bent}. They have been extensively studied for their applications in cryptography, but have also been applied to spread spectrum, coding theory, and combinatorial design.

An indicator, denoted by $Nb_F$, of functions $F$ from an Abelian group (say $A$) to an Abelian group (say $B$) was introduced in 2007 by Carlet and Ding \cite{FFA:CarletD07} as a multiple (to make it an integer) of the variance of the random variable equal to the size of the preimage of a generic element of $B$. No name was given in \cite{FFA:CarletD07} for this parameter. We shall call it the {\em imbalance} of $F$; by definition, it is null for balanced functions (such that all pre-images have the same size) and maximal for constant functions. A deduced parameter $NB_F$ was further introduced for quantifying a kind of nonlinearity  as well, but related to the resistance to the differential attack, equal to the sum of the values of parameter $Nb$ for the derivatives of the function in all nonzero directions. We shall call it the {\em derivative imbalance} of $F$. These two parameters have been studied in \cite{FFA:CarletD07} in relationship with the general nonlinearity parameter $P_F$ introduced in \cite{JC:CarletD04}. Lower and upper bounds, and characterizations of the cases of equality have been derived. In the case of functions from $\GF{}^n$ to $\GF{}^m$ (that are potentially usable as S-boxes) were provided a characterization by means of the fourth moment of the Walsh transform and derived several bounds relating $NB_F$ and the nonlinearity $\mathcal{NL}(F)$; in particular, an upper bound which unifies Sidelnikov-Chabaud-Vaudenay's bound and the covering radius bound was obtained. More results have been shown in \cite{DCC:Carlet11}, in particular on the invariance of these parameters under EA equivalence and CCZ equivalence, and on the characterization of perfect nonlinear (PN) and almost perfect nonlinear (APN) functions (which are important because they are resistant to differential cryptanalysis and for some of them to linear cryptanalysis). Parameter $Nb$ was also made a nonlinearity parameter by considering all the values of $Nb_{F+L}$ where $L$ is any linear function and taking their maximum. Bounds were derived for this nonlinearity parameter. It was also shown that the mean of $L\mapsto Nb_{F+L}$ is the same for all functions, but its variance is directly related to $NB_F$ and depends then on $F$.

In a frequency-hopping radar or sonar system, the signal consists of one or more frequencies, chosen from a set $\{f_1, \ldots, f_n\}$, for transmission at each time interval in a set $\{t_1, \ldots, t_n\}$ of consecutive intervals. Such a signal is conveniently represented by an $n \times n$ permutation matrix $A$, where the $n$ rows correspond to the $n$ frequencies, the $n$ columns correspond to the $n$ time intervals, and the entry $a$, equals $1$ if and only if frequency $f_i$ is transmitted in time interval $t_j$ (otherwise, $a_{ij} = 0$.)
The two-dimensional autocorrelation function $C(r,s)$, called the ambiguity function in the radar and sonar literature, should be thought of as the global ``coincidence'' between the actual returning noisy signal and the shift of the transmitted signal by $r$ units in time and $s$ units in frequency. Costas arrays were first considered by Costas \cite{IEEE:Costas84} as $n\times n$ permutation matrices with ambiguity functions taking only the values 0 and (possibly) 1, and were applied to the processing of radar and sonar signals. A Costas array can also be viewed as a permutation, say $F$, such that each row of the difference triangle (listing the output differences, given a nonzero input difference) contains distinct entries.  The injectivity of $D_a F$ reduces the ambiguity of locating a time and frequency shifted echo of the original signal. Similarly for maps between Abelian groups of the same cardinality, a function $F$ is perfect nonlinear if $D_a F$ is injective and almost perfect nonlinear (APN) if $D_a F$ is at worst 2 to 1. Motivated by the study of Costas arrays and these special functions, a parameter called {\em ambiguity} and denoted by $\mathcal{A}(F)$ of a given bijective mapping $F$ on a finite Abelian group $G$ was introduced in 2010 by Panario \emph{et al.} \cite{LATIN:PanarioSW10,TIT:PanarioSSW11,TIT:PanarioSSTW13} to measure the injectivity of the derivatives $D_a F:G\rightarrow G$ for all nonzero $a\in G$ (for a general function, it measures the imbalance of the derivatives). A second (less important) parameter called {\em deficiency} and denoted by $\mathcal{D}(F)$ was also introduced to measure the surjectivity of the derivatives. The lower the row-$a$ ambiguity of $F$, the closer to be injective is the derivative $D_a F$. Similarly, the lower the row-$a$ deficiency, the closer to be surjective is the derivative $D_a F$. Fundamental results on the ambiguity and deficiency of functions such as their optimality, CCZ-equivalence, as well as the connection with nonlinearity were studied.

Although the motivations of the two indicators $NB_F$ and ambiguity $\mathcal{A}(F)$ are different, {\em we shall see that ambiguity introduced by Panario et al. in 2010 is in fact equivalent to $NB_F$ introduced in 2007 by Carlet and Ding}, up to additive and multiplicative constants, that is, up to a rescaling. It is then necessary to make a comparison between the results of \cite{FFA:CarletD07, DCC:Carlet11} and those of \cite{LATIN:PanarioSW10,TIT:PanarioSSW11,TIT:PanarioSSTW13}, to see which ones are equivalent and which ones are not, and to state the unified versions of these results. We also generalize to any map between any two finite Abelian groups $G_1$ and $G_2$ with possible different orders some results obtained in \cite{FFA:CarletD07, DCC:Carlet11} for  vectorial Boolean functions and in \cite{LATIN:PanarioSW10,TIT:PanarioSSW11,TIT:PanarioSSTW13} for permutations. We systematically investigate these two parameters for general maps between any two finite Abelian groups $G_1$ and $G_2$, including their lower bounds and connections with nonlinearity, Fourier transforms, second order derivatives, among others. Several new results are obtained.

The rest of this paper is organized as follows. In the next section, we recall the definitions and some known results  of  these two  parameters and  explain why the parameter ambiguity is equivalent to the indicator $NB_F$. In Section \ref{sec:ambiguity}, we compare the results on derivative imbalance and those on ambiguity and we generalize several of them to functions between any two finite Abelian groups with possible different orders. A generalization of the characterization of these parameters obtained in \cite{DCC:Carlet11} in terms of the fourth moment of their Fourier transform is also given. Then we further discuss some connections between these parameters, the second-order derivative (which was also investigated in \cite{DCC:Carlet11}) and the autocorrelation of a function. In Section \ref{sec:finitefields}, we give lower bounds of these parameters for arbitrary map $F$, as well as for the functions with differential uniformity $k$. Then we consider maps over the finite fields with characteristic 2, and further results are presented. We also obtain some explicit relations between deficiency and ambiguity for functions with at most $3$ values in their differential spectrum. Conclusions and some open problems are given in Section \ref{sec:conclusion}.

\section{Derivative Imbalabnce and Ambiguity of Mappings Between Abelian Groups}\label{sec:preliminaries}

Let $G_1$ and $G_2$ be two finite Abelian groups (written additively) with orders $\order{G_1}$ and $\order{G_2}$, respectively. We denote by $0$ the identity (neutral) element of an Abelian group. Let $G_1^*=G_1\setminus\{0\}$, and $G_2^*=G_1\setminus\{0\}$. For a function $F:G_1\rightarrow G_2$, the \emph{(first-order) derivative} of $F$ with respect to $a\in G_1$ is defined as
\[
D_a F: x\in G_1 \mapsto F(x+a)-F(x).
\]
The \emph{second-order derivative} of $F$ with respect to $a\in G_1$, $b\in G_1$ is defined as
\[
D_a D_b F: x\in G_1 \mapsto F(x+a+b)-F(x+a)-F(x+b)+F(x).
\]
One can readily see that $D_b D_a F(x)=D_a D_b F(x)$ for all $x\in G_1$.

For any $a\in G_1$ and $b\in G_2$, we define
\[
  \delta_F(a,b)=\order{\{x\in G_1: D_a F(x)=b\}}.
\]
The maximum
\[
  \Delta_F=\max_{a\in G_1^*,b\in G_2}\delta_F(a,b)
\]
is called the \emph{differential uniformity} of $F$. The function $F$ is said to be \emph{almost perfect nonlinear} (APN) if $\Delta_F\leq 2$.

We denote by $N_i$ the number of pairs of nonzero input difference $a$ and output difference $b$ that occur $i$ times
\[
N_i = \order{\{(a,b)\in G_1^*\times G_2 : \delta_F(a,b)=i\}}.
\]
The \emph{differential spectrum} of $F$ is the set:
\[
\mathfrak{D}_F=\{N_0,N_1,\dots,N_{\Delta_F}\}.
\]
Obviously, the differential spectrum of $F$ satisfies
\begin{equation}\label{eqn:sumofN_i}
\sum_{i=0}^{\Delta_F}N_i=(\order{G_1}-1)\order{G_2}
\end{equation}
and
\begin{equation}\label{eqn:sumofiN_i}
\sum_{i=0}^{\Delta_F}i\times N_i=(\order{G_1}-1)\order{G_1}.
\end{equation}

\subsection{Derivative Imbalance and Ambiguity}

When  two finite Abelian groups $G_1$ and $G_2$ have different orders, bijections between them of course do not exist and the proper concept  generalizing the notion of bijectivity is \emph{balancedness}. A function is balanced if, for each $b\in G_2$, it holds that
\[
  \order{\{x\in G_1:F(x)=b\}}=\frac{\order{G_1}}{\order{G_2}}.
\]
By definition, a necessary condition for the existence of balanced functions from $G_1$ to $G_2$ is that $\order{G_2}$  be a factor of $\order{G_1}$. A function $F$ from $G_1$ to $G_2$ is \emph{perfect nonlinear} if and only if all of its nonzero derivatives are balanced in $G_1$.

Next we give the definitions of imbalance and derivative imbalance of a function, which were introduced in 2007 by Carlet and Ding \cite{FFA:CarletD07} (without that names be given) for quantifying the balancedness of a function, respectively, of its derivatives. The imbalance is the variance of the random variable $b\mapsto \order{F^{-1}(b)}$ (where $b$ ranges uniformly over $G_2$), multiplied by the factor $\order{G_2}$, so that it be an integer when $\order{G_2}$ divides $\order{G_1}$:
\begin{defi}[See \cite{FFA:CarletD07}]
  Let $F$ be a function from $G_1$ to $G_2$. Then the \emph{imbalance} of $F$ is defined as
  \[
   Nb_F=\sum_{b\in G_2}\left(\order{F^{-1}(b)}-\frac{\order{G_1}}{\order{G_2}}\right)^2=\sum_{b\in G_2}\order{F^{-1}(b)}^2-\frac{\order{G_1}^2}{\order{G_2}},
  \]
  and the \emph{derivative imbalance} of $F$ is defined as
  \[
   NB_F = \sum_{a \in G_1^*} Nb_{D_a F}.
  \]
\end{defi}

Note that
\[
  NB_F = \sum_{a \in G_1^*}\sum_{b\in G_2}\order{(D_aF)^{-1}(b)}^2-\left(\order{G_1}-1\right)\frac{\order{G_1}^2}{\order{G_2}}.
\]

Independently, motivated by the study of Costas arrays, ambiguity and deficiency of a bijective mapping  are introduced in \cite{LATIN:PanarioSW10}:
\begin{defi}[See \cite{LATIN:PanarioSW10}]
  Let $F$ be a function from $G_1$ to $G_2$. Then the \emph{ambiguity} of $F$ is defined as
  \[
  \mathcal{A}(F)=\sum_{i=0}^{\Delta_F}N_i\binom{i}{2},
  \]
  and the \emph{deficiency} of $F$ is defined as
  \[
  \mathcal{D}(F)=N_0.
  \]
\end{defi}

By definition, it is easy to see that the ambiguity is equal to the total replication number of pairs of $x$ and $y$ such that $D_a F(x)=D_a F(y)$ for all $a\in G_1^*$. The deficiency is equal to the number of pairs $(a,b)\in G_1^*\times G_2$ such that $D_a F(x)=b$ has no solution. These parameters are closely related to the balancedness of the derivatives, since it is easy to see that if a function has its output distributions of all derivatives close to the uniform distribution, then ambiguity and deficiency are as low as possible.

\subsection{Ambiguity as a Rescaling of Derivative Imbalance}

In fact, ambiguity can be determined completely from $NB_F$. We have:
\begin{align}
  \mathcal{A}(F) & = \sum_{i=0}^{\Delta_F}N_i\binom{i}{2}=\sum_{a\in G_1^*}\sum_{b\in G_2}\binom{\order{(D_aF)^{-1}(b)}}{2}\notag \\
                 & = \frac{1}{2}\sum_{a\in G_1^*}\sum_{b\in G_2}\order{(D_aF)^{-1}(b)}^2-\frac{1}{2}\sum_{a\in G_1^*}\sum_{b\in G_2}\order{(D_aF)^{-1}(b)} \notag \\
                 & = \frac{1}{2}\left(NB_F+\left(\order{G_1}-1\right)\frac{\order{G_1}^2}{\order{G_2}}\right)-\frac{1}{2}\left(\order{G_1}-1\right)\order{G_1}. \notag \\
                 & = \frac{1}{2}NB_F+\frac{1}{2}\left(\order{G_1}-1\right)\left(\frac{\order{G_1}^2}{\order{G_2}}-\order{G_1}\right). \label{eqn:nbambiguity}
\end{align}
Therefore, the two indicators are equivalent up to additive and multiplicative constants. One can obtain the indicator $\mathcal{A}(F)$ from the indicator $NB_F$, and vice versa. In particular, if
$\order{G_1} = \order{G_2}$ then $\mathcal{A}(F) = \frac{1}{2} NB_F$.

\subsection{Linearity and Nonlinearity}

Given a complex number $z\in\mathbb{C}$, $|z|$ and $\overline{z}$ denote the absolute value and the conjugate of $z$, respectively. Let $G$ be a finite Abelian group. The \emph{Fourier transform} of any complex-value function $\Phi$ on $G$ is defined by
\[
\widehat{\Phi}(\chi)=\sum_{x\in G}\Phi(x)\overline{\chi(x)},
\]
where $\chi$ is a character of $G$. It is well known that the characters of $G$ form a group $G$\^{} isomorphic to $G$. Denoting by $\chi_{\alpha}$ the image of $\alpha\in G$ under an arbitrary but fixed isomorphism from $G$ to $G$\^{}, then we can write this as
\[
\widehat{\Phi}(\alpha)=\sum_{x\in G}\Phi(x)\overline{\chi_{\alpha}(x)}.
\]
As a result, we can consider $\widehat{\Phi}$ to be defined on the group $G$.

Now we consider the function $F$ between two finite Abelian groups $G_1$ and $G_2$. Again identifying $\psi_{\beta}$ as the image of $\beta$ under an arbitrary but fixed isomorphism from $G_2$ to $G_2$\^{}, then we define the \emph{Fourier transform} of $F$ at $\alpha\in G_1$ and $\beta\in G_2$ by
\[
\widehat{F}(\alpha,\beta)=\sum_{x\in G_1}\psi_{\beta}(F(x))\overline{\chi_{\alpha}(x)}.
\]

The linearity of $F$ is studied through the Fourier transform and is then given by the following definition.
\begin{defi}[See \cite{DAM:Pott04,TIT:DrakakisRM10}]\label{def:linearity}
  Let $F$ be a function from $G_1$ to $G_2$, the \emph{linearity} of $F$ is defined by
  \[
  \mathcal{L}(F)=\max_{\alpha\in G_1, \beta\in G_2^*}|\widehat{F}(\alpha,\beta)|.
  \]
  The corresponding \emph{nonlinearity} is given by the following normalized measure.
  \[
  \mathcal{NL}(F)=\frac{\order{G_1}-\mathcal{L}(F)}{\order{G_2}}.
  \]
\end{defi}

It is noticed that the definition \ref{def:linearity} of nonlinearity is normalized, which is different from the classical definition of nonlinearity for a function from $\GF{n}$ to $\GF{m}$ when $m>1$. For the sake of comparison with known results, when considering the nonlinearity of a function between finite fields with characteristic 2, we always refer to the classic definition $\mathcal{NL}(F)=2^{n-1}-\frac{1}{2}\mathcal{L}(F)$.

The following orthogonality relations for characters are well known:
\begin{lem}[See \cite{FFields:LidlN97}]
Let $G$ be a finite Abelian group with identity $0$, then the following two identities hold:
\[
\sum_{x\in G}\chi(x)=
\begin{cases}
  0, & \mbox{if } \chi\ne\chi_0 \\
  \order{G}, & \mbox{otherwise},
\end{cases}
\]
and
\[
\sum_{\chi\in G\text{\^{}}}\chi(x)=
\begin{cases}
  0, & \mbox{if } x\ne 0 \\
  \order{G}, & \mbox{otherwise}.
\end{cases}
\]
\end{lem}

Assume that $F$ is a function from $G_1$ to $G_2$, then we have that for any $\alpha\in G_1$ and $\beta\in G_2^*$,
\begin{equation}\label{equ:squarefourier}
\begin{split}
    |\widehat{F}(\alpha,\beta)|^2 =  &\left(\sum_{x\in G_1}\psi_{\beta}(F(x))\overline{\chi_{\alpha}(x)}\right)\left(\overline{\sum_{y\in G_1}\psi_{\beta}(F(y))\overline{\chi_{\alpha}(y)}}\right) \\
 =  &\sum_{x\in G_1}\sum_{y\in G_1}\psi_{\beta}(F(x)-F(y))\chi_{\alpha}(y-x) \\
 =  &\sum_{x\in G_1}\sum_{a\in G_1}\psi_{\beta}(-D_a F(x))\chi_{\alpha}(a). \\
\end{split}
\end{equation}
We still have the following Parseval's relation
\begin{align*}
\sum_{\alpha\in G_1}|\widehat{F}(\alpha,\beta)|^2 & = \sum_{\alpha\in G_1}\sum_{x\in G_1}\sum_{a\in G_1}\psi_{\beta}(-D_a F(x))\chi_{\alpha}(a) \\
                                                  & = \sum_{x\in G_1}\sum_{a\in G_1}\psi_{\beta}(-D_a F(x))\left(\sum_{\alpha\in G_1}\chi_{\alpha}(a)\right) \\
                                                  & = \order{G_1}\sum_{x\in G_1}\psi_{\beta}(-D_0 F(x)) \\
                                                  & = \order{G_1}^2.
\end{align*}

Hence, $\max_{\alpha\in G_1,\beta\in G_2^*}|\widehat{F}(\alpha,\beta)|\geq \sqrt{\order{G_1}}$. It is trivial that $|\widehat{F}(\alpha,\beta)|\leq\order{G_1}$. So we have obtained that
\[
\sqrt{\order{G_1}} \leq \mathcal{L}(F) \leq \order{G_1}.
\]
Functions with $\mathcal{L}(F)=\sqrt{\order{G_1}}$ are called perfect nonlinear.

Autocorrelation is a measure of the proximity between a function and its shift. It is a useful tool to characterize the differential uniformity of a function. In the following we introduce the definition of autocorrelation functions between any two finite Abelian groups.
\begin{defi}
  Let $F$ be a function from $G_1$ to $G_2$, then the \emph{autocorrelation function} of $F$ at $\alpha\in G_1$ and $\beta\in G_2$ is defined as
  \[
  \mathcal{C}_F(\alpha,\beta)=\sum_{x\in G_1}\psi_{\beta}(F(x+\alpha))\overline{\psi_{\beta}(F(x))}.
  \]
\end{defi}

\subsection{EA-equivalence and CCZ-equivalence}

In the classical case of S-boxes, EA-equivalence and CCZ-equivalence are two relevant notions of equivalence with respect to the differential and linearity properties of a function since they preserve both the differential and the Fourier spectra. Next we give the general definitions of these two kinds of equivalence, which were introduced in \cite{TIT:PanarioSSTW13}.

A function $L:G_1 \mapsto G_2$ is \emph{linear} if $L(x+y)=L(x)+L(y)$ for all $x,y\in G_1$. A function $A:G_1 \mapsto G_2$ is \emph{affine} if $A(x+y)=A(x)+A(y)+c$ for a fixed constant $c\in G_2$ and all $x,y\in G_1$.

Let $G_1$ and $G_2$ be arbitrary groups. Two functions $F_1$ and $F_2: G_1 \mapsto G_2$ are called \emph{extended affine equivalent} (EA-equivalent), if there exist affine permutations $A_1: G_1\mapsto G_1$, $A_2: G_2\mapsto G_2$ and an affine function $A_3:G_1 \mapsto G_2$ such that $F_2=A_2\circ F_1 \circ A_1 +A_3$. In particular, if $A_3=0$, then $F_1$ and $F_2$ are called \emph{affine equivalent}.

Two functions $F_1$ and $F_2: G_1 \mapsto G_2$ are called \emph{Carlet-Charpin-Zinoviev equivalent} (CCZ-equivalent), if there exists an affine permutation $A: G_1\times G_2\mapsto G_1\times G_2$ that maps $\mathcal{G}_{F_1}$ to $\mathcal{G}_{F_2}$, where $\mathcal{G}_{F_1}=\{(x,F_1(x))\in G_1\times G_2 :x\in G_1\}$ is the \emph{graph} of $F_1$, and $\mathcal{G}_{F_2}=\{(x,F_2(x))\in G_1\times G_2 :x\in G_1\}$  is the \emph{graph} of $F_2$.

Similarly as the S-boxes, we still have that EA-equivalence implies CCZ-equivalence. It was also shown that the indicators $NB_F$, ambiguity, deficiency, linearity and nonlinearity are invariant under EA-equivalence and CCZ-equivalence \cite{DCC:CarletCZ98,JC:CarletD04,TIT:PanarioSSTW13}.

\section{Known Results and Their Generalizations}\label{sec:ambiguity}

In this section, we recall the main known results on the two indicators, $NB_F$ and $\mathcal{A}(F)$. Thanks to the equivalence relation \eqref{eqn:nbambiguity}, some bounds on one indicator can be obtained or improved from the other indicator. Then we further generalize the characterizations of these parameters obtained in \cite{DCC:Carlet11} in terms of the fourth moment of their Fourier transform and the second-order derivative to functions between any two finite Abelian groups with different orders. We also discuss the connection between these parameters and the autocorrelation function.

\subsection{Known Results}

Firstly, the followings are some basic facts from \cite{FFA:CarletD07}.
\begin{prop}[See \cite{FFA:CarletD07}]\label{prop:NBofCarlet}
Let $F$ be a function from $G_1$ to $G_2$. Then the following statements hold:
\begin{enumerate}
\item $NB_F \geq0$ with equality if and only if $F$ is perfect nonlinear. \label{item:LowBoundNB}
\item If $G$ differs from $F$ by an affine function, then $NB_F=NB_G$.
\item $NB_F \leq (\order{G_1}-1)\left(\order{G_1}^2 - \frac{\order{G_1}^2}{\order{G_2}}\right)$ with equality if and only if $F$ is affine. \label{item:UpperBoundNB}
\item For any affine function $A:G_1\mapsto G_2$, we have
      \[
       NB_F\geq \frac{\order{G_2}}{(\order{G_1}-1)(\order{G_2}-1)}\left(Nb_{F+A}-\left(\order{G_1}-\frac{\order{G_1}}{\order{G_2}}\right)\right)^2.
      \]
\item If we denote $T_F=\max_{0\neq a\in G_1}\order{\mathrm{Im}(D_aF)}$, then
      \begin{equation}\label{eqn:LowerBoundByTF}
       NB_F \geq \order{G_1}^2(\order{G_1}-1)\left(\frac{1}{T_F}-\frac{1}{\order{G_2}}\right).
      \end{equation}
\end{enumerate}
\end{prop}

If the characteristic of $G_1$ is $2$,  then the lower bound \eqref{eqn:LowerBoundByTF} becomes
\begin{equation}\label{eqn:BoundofNB}
NB_{F}\geq \max \left\{0,  (\order{G_1}-1)\left(2\order{G_1} - \frac{\order{G_1}^2}{\order{G_2}} \right) \right\}.
\end{equation}
For $\order{G_1} \leq 2 \order{G_2}$, this bound is achieved by APN functions. In terms of ambiguity, we can restate the above bounds as follows:
\begin{enumerate}
\item  $\mathcal{A}(F)\geq \frac{\order{G_1}-1}{2}\left(\frac{\order{G_1}^2}{\order{G_2}}-\order{G_1}\right)$ with equality if and only if $F$ is perfect nonlinear.
\item  $\mathcal{A}(F)\leq \frac{\order{G_1}-1}{2}\left(\order{G_1}^2-\order{G_1}\right)$ with equality if and only if $F$ is affine.
\item  $\mathcal{A}(F)\geq  \frac{ \order{G_1}\left(\order{G_1}-1\right)}{2} \left(\frac{\order{G_1}}{T_F} -1 \right)$.
\end{enumerate}

In particular, if the characteristic of $G_1$ is $2$, then $\mathcal{A}(F)\geq \max\left\{\frac{\order{G_1}-1}{2}\left(\frac{\order{G_1}^2}{\order{G_2}}-\order{G_1}\right), \frac{\order{G_1}\left(\order{G_1}-1\right)}{2}\right\}$. Furthermore,  the case of S-boxes, namely $G_1=\GF{n}$ and $G_2=\GF{m}$, are intensively studied. Some bounds on the nonlinearity are deduced respectively from the indicator $NB_F$ and coding theory \cite{DCC:Carlet11}.
\begin{prop}[See \cite{DCC:Carlet11}]\label{prop:BoundofNL}
  For any function $F$ from $\GF{n}$ to $\GF{m}$, the following statements hold:
  \begin{enumerate}
    \item
    \begin{equation}\label{eqn:NLBoundNB}
      \mathcal{NL}(F)\leq 2^{n-1}-\frac{1}{2}\sqrt{2^n+\frac{2^{m-n}}{2^m-1}NB_F}.
    \end{equation}
    \item
    \begin{equation}\label{eqn:UnifyingBound}
     \mathcal{NL}(F)\leq 2^{n-1}-\frac{1}{2}\sqrt{\frac{(2^n-1)2^{n+m-\min\{m,n-1\}}+2^{n+m}-2^{2n}}{2^m-1}}.
    \end{equation}
    \item
    When $m<2^n-2$,
    \[
     \mathcal{NL}(F)\leq 2^{n-1}-\frac{m}{2}\times\frac{2^{n-1}}{2^{n-1}-1}.
    \]
    \item
    When $m<2^n-n$,
    \[
     \mathcal{NL}(F)\leq 2^{n-1}-n-m.
    \]
    \item
    \[
     \sum_{i=0}^{\left\lfloor\frac{\mathcal{NL}(F)-1}{2}\right\rfloor}\binom{2^n}{i}\leq 2^{2^n-n-m-1}.
    \]
  \end{enumerate}
\end{prop}
The bound (\ref{eqn:NLBoundNB}) can be restated as
\[
\mathcal{NL}(F)\leq 2^{n-1}-\frac{1}{2}\sqrt{2^n+\frac{2^{m-n+1}}{2^m-1}\mathcal{A}(F) - \frac{2^n-1}{2^m-1} \left(2^n -2^m\right)}.
\]
When $1\leq m\leq n-1$, the bound of \eqref{eqn:UnifyingBound} becomes the covering radius bound. When $m\geq n$, it becomes Sidelnikov-Chabaud-Vaudenay's bound. Thus the bound of \eqref{eqn:UnifyingBound} is an unification of the two bounds.

When $G_1$ and $G_2$ have the same size, the bijections from $G_1$ to $G_2$ are studied frequently for their practical application in encryption and decryption. However, perfect nonlinear functions cannot exist in this case, and the lower bound, $NB_F\geq 0$, in Proposition \ref{prop:NBofCarlet} can never be achieved. In \cite{TIT:PanarioSSW11}, Panario \emph{et al.} investigated the optimum lower bounds of the parameters ambiguity and deficiency in this case, which says,
\begin{prop}[See \cite{TIT:PanarioSSW11}]\label{prop:AmbiguityToNB}
  Let $G_1$ and $G_2$ be two Abelian groups of order $n$ with $\iota_1$ and $\iota_2$ elements of order 2, respectively. Let $F:G_1\rightarrow G_2$ be a bijection. Then
  \[
   \mathcal{A}(F)\geq
   \begin{cases}
     2(n-1), & \mbox{if } n\equiv 1\pmod{2} \\
     2(n-2), & \mbox{if } n\equiv 0\pmod{2} \mbox{ and }\iota_1\iota_2=1 \\
     2(n-1)-\frac{3\min\{\iota_1,\iota_2\}}{2}+\frac{\iota_1\iota_2}{2}, & \mbox{if } n\equiv 0\pmod{2} \mbox{ and }\iota_1\iota_2>1.
   \end{cases}
  \]
\end{prop}

Therefore, the previous proposition gives a nontrivial improvement on the lower bound of indicator $NB_F$ in this case, namely
\[
   NB_F\geq
   \begin{cases}
     4(n-1), & \mbox{if } n\equiv 1\pmod{2} \\
     4(n-2), & \mbox{if } n\equiv 0\pmod{2} \mbox{ and }\iota_1\iota_2=1 \\
     4(n-1)-3\min\{\iota_1,\iota_2\}+\iota_1\iota_2, & \mbox{if } n\equiv 0\pmod{2} \mbox{ and }\iota_1\iota_2>1.
   \end{cases}
\]
In general, this lower bound is sharp. A bijection from $G_1$ to $G_2$ whose ambiguity achieves this lower bound is said to have \emph{optimum ambiguity}. In \cite{TIT:PanarioSSW11}, Panario \emph{et al.} also obtained several constructions which have optimum ambiguity or nearly optimum ambiguity in the cyclic group $\mathbb{Z}_n$ where $n=p^m-1$ and $p$ is a prime number. Furthermore, in \cite{TIT:PanarioSSTW13}, Panario \emph{et al.} investigated the lower bound on the nonlinearity of permutations which achieve optimum ambiguity.

For the special case $G_1=G_2=G$, the lower and upper bounds on the ambiguity of differentially $k$-uniform functions are also provided in \cite{TIT:PanarioSSTW13}.
\begin{prop}[See \cite{TIT:PanarioSSTW13}]\label{prop:AmbiguityOfDiffK}
  Let $F:G\rightarrow G$ be a function with differential uniformity $k$. Suppose further that $\order{G}=n=rk+s$, for some integers $r$, $s$ with $0\leq s<k$. Then the ambiguity of $F$ satisfies
  \[
    \binom{k}{2}\leq \mathcal{A}(F)\leq (n-1)\left(r\binom{k}{2}+\binom{s}{2}\right).
  \]
\end{prop}

In terms of derivative imbalance $NB_F$ and with the same notation, Proposition \ref{prop:AmbiguityOfDiffK} says that,
\[
  k(k-1) \leq NB_F \leq (n-1)\left(rk^2+s^2-n\right).
\]
Note that when $k\ne n$ (in other words, $F$ is not an affine function), we always have that
\[
  rk^2+s^2<n^2=(rk+s)^2=r^2k^2+s^2+2rks.
\]
Hence, Proposition \ref{prop:AmbiguityOfDiffK} derives a nontrivial refinement on the upper bound $NB_F\leq (n-1)(n^2-n)$ that presented in Proposition \ref{prop:NBofCarlet}.

\subsection{The Generalizations of Some Known Results}

In the sequel we aim to systematically investigate these two parameters and their connections with nonlinearity, Fourier transforms, second order derivatives, among others. Since most functions considered previously for these results are vectorial Boolean functions in \cite{FFA:CarletD07,DCC:Carlet11} and permutations in \cite{LATIN:PanarioSW10, TIT:PanarioSSW11,TIT:PanarioSSTW13}, our study deals with the more general situation where $F$ is any map between any two finite Abelian groups $G_1$ and $G_2$ with possible different orders.

In the case of $G_1=\GF{n}$, $G_2=\GF{m}$, it is shown in \cite{DCC:Carlet11} that
\begin{align*}
  NB_F & = \sum_{a\in\GF{n}^*}\order{\{(x,y)\in\GF{n}\times\GF{n}:D_aF(x)=D_aF(y)\}}-(2^n-1)2^{2n-m} \\
       & = \order{\left\{(x,x',y,y')\in(\GF{n})^4:\begin{cases}x+x'=y+y'\neq 0 \\ F(x)+F(x')=F(y)+F(y')\end{cases}\right\}}-(2^n-1)2^{2n-m}.
\end{align*}
The following corollary is an analogue of the above result, which comes directly from the observation below the definition of ambiguity. We shall use it frequently later.
\begin{cor}\label{cor:ambiguity}
  Let $F$ be a function from $G_1$ to $G_2$. Then
  \[
   NB_F= \sum_{a\in G_1}\order{\{(x,y)\in G_1\times G_1:D_a F(x)=D_a F(y)\}}-\order{G_1}^2-\left(\order{G_1}-1\right)\frac{\order{G_1}^2}{\order{G_2}},
  \]
  or equivalently,
  \[
  \mathcal{A}(F)= \frac{1}{2}\sum_{a\in G_1}\order{\{(x,y)\in G_1\times G_1:D_a F(x)=D_a F(y)\}}-\order{G_1}^2+\frac{\order{G_1}}{2}.
  \]
\end{cor}

In \cite{FFA:CarletD07}, when $G_1=\GF{n}$ and $G_2=\GF{m}$, Carlet and Ding showed that
\[
\sum_{u\in\GF{n}}\sum_{v\in\GF{m}^*}|\widehat{F}(u,v)|^4=2^{3n}(2^m-1)+2^{n+m}NB_F.
\]
It is easy to generalize this equality to functions defined over any two finite Abelian groups. Next we derive a characterization on these parameters by means of the fourth moment of its Fourier transform. Then by \eqref{equ:squarefourier}, we have
\begin{align*}
  \sum_{\alpha\in G_1}\sum_{\beta\in G_2}|\widehat{F}(\alpha,\beta)|^4 = & \sum_{\alpha\in G_1}\sum_{\beta\in G_2}\left(\sum_{x\in G_1}\sum_{a\in G_1}\psi_{\beta}(-D_a F(x))\chi_{\alpha}(a)\right) \\
    & \phantom{\sum_{\alpha\in G_1}\sum_{\beta\in G_2}}\cdot\left(\overline{\sum_{y\in G_1}\sum_{b\in G_1}\psi_{\beta}(-D_b F(y))\chi_{\alpha}(b)}\right) \\
  = & \sum_{\alpha\in G_1}\sum_{\beta\in G_2}\sum_{x,y,a,b\in G_1}\psi_{\beta}(D_b F(y)-D_a F(x))\chi_{\alpha}(a-b) \\
  = & \sum_{x,y\in G_1}\sum_{a,b\in G_1}\left(\sum_{\beta\in G_2}\psi_{\beta}(D_b F(y)-D_a F(x))\right)\left(\sum_{\alpha\in G_1}\chi_{\alpha}(a-b)\right) \\
  = & \order{G_1}\sum_{x,y\in G_1}\sum_{a\in G_1}\left(\sum_{\beta\in G_2}\psi_{\beta}(D_a F(y)-D_a F(x))\right) \\
  = & \order{G_1}\order{G_2}\sum_{a\in G_1}\order{\{(x,y)\in G_1\times G_1:D_a F(x)=D_a F(y)\}}.
\end{align*}
By Corollary \ref{cor:ambiguity}, we deduce that
\begin{prop}\label{prop:4thfourier}
  Assume that $G_1$ and $G_2$ are two finite Abelian groups. Let $F$ be a function from $G_1$ to $G_2$. Then
  \[
   NB_F= \frac{1}{\order{G_1}\order{G_2}}\sum_{\alpha\in G_1}\sum_{\beta\in G_2}|\widehat{F}(\alpha,\beta)|^4-\order{G_1}^2-\left(\order{G_1}-1\right)\frac{\order{G_1}^2}{\order{G_2}},
  \]
  or equivalently,
  \[
  \mathcal{A}(F)=\frac{1}{2\order{G_1}\order{G_2}}\sum_{\alpha\in G_1}\sum_{\beta\in G_2}|\widehat{F}(\alpha,\beta)|^4-\order{G_1}^2+\frac{\order{G_1}}{2}.
  \]
\end{prop}

By the Cauchy-Schwartz inequality and Parseval's relation:
\begin{equation}\label{bnd:ntomfunctions1}
\begin{split}
\sum_{\beta\in G_2}\sum_{\alpha\in G_1}|\widehat{F}(\alpha,\beta)|^4 & = \sum_{\alpha\in G_1}|\widehat{F}(\alpha,0)|^4+\sum_{\beta\in G_2^*}\sum_{\alpha\in G_1}|\widehat{F}(\alpha,\beta)|^4 \\
& \geq \order{G_1}^4+\sum_{\beta\in G_2^*}\frac{\left(\sum_{\alpha\in G_1}|\widehat{F}(\alpha,\beta)|^2\right)^2}{\order{G_1}} \\
& = \order{G_1}^4+(\order{G_2}-1)\order{G_1}^3
\end{split}
\end{equation}
and then by Propposition \ref{prop:4thfourier}, we have $NB_F\geq 0$, which gives another proof of the bound given in Proposition \ref{prop:NBofCarlet}.

The characterization in Proposition \ref{prop:4thfourier} gives a nontrivial upper bound on the nonlinearity of a function by its derivative imbalance or ambiguity. This is an analogue of bound \eqref{eqn:NLBoundNB} from Proposition \ref{prop:BoundofNL}.
\begin{cor}\label{cor:AmbiguityNonlinearity}
  Assume that $G_1$ and $G_2$ are two finite Abelian groups. Let $F$ be a function from $G_1$ to $G_2$. Then
  \[
   \mathcal{NL}(F)\leq \frac{\order{G_1}}{\order{G_2}}-\frac{1}{\order{G_2}}\sqrt{\order{G_1}+\frac{\order{G_2}}{\order{G_1}(\order{G_2}-1)}NB_F},
  \]
  or equivalently,
  \[
   \mathcal{NL}(F)\leq \frac{\order{G_1}}{\order{G_2}}-\frac{1}{\order{G_2}}\sqrt{\order{G_1}+\frac{2\order{G_2}}{\order{G_1}(\order{G_2}-1)}\mathcal{A}(F) -\frac{\order{G_1}-1}{\order{G_2}-1}\left(\order{G_1}-\order{G_2}\right) }
  \]
\end{cor}

\begin{rmk}
If $F$ is the Gold function $x^{2^i+1}$ over $\GF{2k}$, where $k$ is odd and $\gcd(i,2k)=1$, it is well known that all the nonzero derivatives $D_aF$ are 4-to-1. Hence, we have $\mathcal{A}(F)=3\cdot 2^{2k-1}(2^{2k}-1)$. The bound in Corollary \ref{cor:AmbiguityNonlinearity} is thus achieved because the Gold function has the best known nonlinearity $2^{2k-1}-2^k$. Note here we refer to the definition $\mathcal{NL}(F)=2^{n-1}-\frac{1}{2}\mathcal{L}(F)$ over finite fields $\GF{n}$.
\end{rmk}

With respect to the autocorrelation function, By definition, we have
\[
\mathcal{C}_F(\alpha,\beta)=\sum_{x\in G_1}\psi_{\beta}(D_{\alpha}F(x)).
\]
Then,
\begin{align*}
  \sum_{\alpha\in G_1}\sum_{\beta\in G_2}|\mathcal{C}_F(\alpha,\beta)|^2 = & \sum_{\alpha\in G_1}\sum_{\beta\in G_2}\sum_{x\in G_1}\psi_{\beta}(D_{\alpha}F(x))\overline{\sum_{y\in G_1}\psi_{\beta}(D_{\alpha}F(y))} \\
  = & \sum_{\alpha\in G_1}\sum_{x\in G_1}\sum_{y\in G_1}\left(\sum_{\beta\in G_2}\psi_{\beta}(D_{\alpha}F(x)-D_{\alpha}F(y))\right) \\
  = & \order{G_2}\sum_{\alpha\in G_1}\order{\{(x,y)\in G_1\times G_1:D_{\alpha}F(x)=D_{\alpha}F(y)\}}.
\end{align*}
Combining with Corollary \ref{cor:ambiguity}, we give another characterization by the autocorrelation functions.

\begin{prop}\label{prop:autocorrelation}
  Assume that $G_1$ and $G_2$ are two finite Abelian groups. Let $F$ be a function from $G_1$ to $G_2$. Then
  \[
    NB_F= \frac{1}{\order{G_2}}\sum_{\alpha\in G_1}\sum_{\beta\in G_2}|\mathcal{C}_F(\alpha,\beta)|^2-\order{G_1}^2-\left(\order{G_1}-1\right)\frac{\order{G_1}^2}{\order{G_2}},
  \]
  or equivalently,
  \[
   \mathcal{A}(F)=\frac{1}{2\order{G_2}}\sum_{\alpha\in G_1}\sum_{\beta\in G_2}|\mathcal{C}_F(\alpha,\beta)|^2-\order{G_1}^2+\frac{\order{G_1}}{2}.
  \]
\end{prop}

Let us now study the connection between these two parameters and its second-order derivative of a function. Firstly, we recall the following result given by Carlet \cite{DCC:Carlet11} for the case of vectorial Boolean functions, namely for any function from $\GF{n}$ to $\GF{m}$:
\[
  NB_F=\sum_{\substack{a,a'\in\GF{n}\\ \text{linearly indept}}}\order{(D_aD_{a'})^{-1}(0)}-(2^n-1)(2^{2n-m}-2^{n+1}).
\]
Similarly, for any function from $G_1$ to $G_2$,
\begin{align*}
  \sum_{\beta\in G_2}\sum_{a,b,x\in G_1}\psi_{\beta}(D_a D_b F(x)) = & \sum_{a,b,x\in G_1}\left(\sum_{\beta\in G_2}\psi_{\beta}(D_a D_b F(x))\right) \\
  = & \order{G_2}\sum_{a\in G_1}\order{\{(b,x)\in G_1\times G_1:D_a D_b F(x)=0\}} \\
  = & \order{G_2}\sum_{a\in G_1}\order{\{(b,x)\in G_1\times G_1:D_a F(x+b)=D_a F(x)\}} \\
  = & \order{G_2}\sum_{a\in G_1}\order{\{(y,x)\in G_1\times G_1:D_a F(y)=D_a F(x)\}}.
\end{align*}
Therefore this derives another characterization by the second-derivative, and we state it in the following proposition.

\begin{prop}\label{prop:2ndderivative}
  Assume that $G_1$ and $G_2$ are two finite Abelian groups. Let $F$ be a function from $G_1$ to $G_2$. Then
  \[
   NB_F= \frac{1}{\order{G_2}}\sum_{\beta\in G_2}\sum_{a,b,x\in G_1}\psi_{\beta}(D_a D_b F(x))-\order{G_1}^2-\left(\order{G_1}-1\right)\frac{\order{G_1}^2}{\order{G_2}},
  \]
  or equivalently,
  \[
  \mathcal{A}(F)=\frac{1}{2\order{G_2}}\sum_{\beta\in G_2}\sum_{a,b,x\in G_1}\psi_{\beta}(D_a D_b F(x))-\order{G_1}^2+\frac{\order{G_1}}{2}.
  \]
\end{prop}

For example, let us consider the monomial $x^{p^i+p^j}$ over $\mathbb{F}_{p^n}$ where $p$ is odd (see Example \ref{exa:PowerPlateaued} for even $p$) and $i\geq j$. For any $a, b\in\mathbb{F}_{p^n}$, the second-derivative is $D_aD_b F(x)=(x+a+b)^{p^i+p^j}-(x+a)^{p^i+p^j}-(x+b)^{p^i+p^j}+x^{p^i+p^j}=a^{p^i}b^{p^j}+a^{p^j}b^{p^i}=D_aD_bF(0)$, which is bilinear with respect to $a$ and $b$. Then we have
\begin{align*}
NB_F           & = \frac{1}{p^n}\sum_{\beta\in\mathbb{F}_{p^n}}\sum_{a,b,x\in\mathbb{F}_{p^n}}\psi_{\beta}(D_a D_b F(0))-p^{2n}-p^n(p^n-1)  \\
               & = \sum_{a,b\in\mathbb{F}_{p^n}}\sum_{\beta\in\mathbb{F}_{p^n}}\psi_{\beta}(D_a D_b F(0))-2p^{2n}+p^n  \\
               & = p^n\order{\{(a,b)\in\mathbb{F}_{p^n}\times\mathbb{F}_{p^n}:a^{p^i}b^{p^j}+a^{p^j}b^{p^i}=0\}}-2p^{2n}+p^n \\
               & = p^n\sum_{a\in\mathbb{F}_{p^n}^*}\order{\{b\in\mathbb{F}_{p^n}:a^{p^i}b^{p^j}+a^{p^j}b^{p^i}=0\}}-p^{2n}+p^n.
\end{align*}

Let $\gamma$ be a primitive element of $\mathbb{F}_{p^n}$ and $s=\gcd(i-j,n)$. If $a\neq 0$, then the equation $a^{p^i}x^{p^j}+a^{p^j}x^{p^i}=0$ is equivalent to that $\left(\left(\frac{x}{a}\right)^{p^{i-j}}+\frac{x}{a}\right)^{p^j}=0$. It is easy to see that the nonzero solutions satisfy $\left(\frac{x}{a}\right)^{p^{i-j}-1}=-1$. Note that $p$ is odd, thus $\frac{x}{a}\in\langle\gamma^{\frac{p^n-1}{\gcd(2(p^{i-j}-1),p^n-1)}}\rangle\setminus\langle\gamma^{\frac{p^n-1}{\gcd(p^{i-j}-1,p^n-1)}}\rangle$. The number of solutions of equation $a^{p^i}x^{p^j}+a^{p^j}x^{p^i}=0$ in $\mathbb{F}_{p^n}$ is equal to $\gcd(2(p^{i-j}-1),p^n-1)-\gcd(p^{i-j}-1,p^n-1)+1$. This is equal to $p^s$ if $\frac{n}{s}$ is even, and equal to 1 otherwise. Therefore,
\[
NB_F=
\begin{cases}
  p^n(p^n-1)(p^s-1), & \mbox{if } \frac{n}{s} \mbox{ is even} \\
  0,                 & \mbox{otherwise}.
\end{cases}
\]
Actually, when $\frac{n}{\gcd(i-j,n)}$ is odd, the monomial $x^{p^i+p^j}$ is perfect nonlinear (or planar) \cite{MZ:DembowskiO68}. Moreover, for any DO polynomial over $\mathbb{F}_{p^n}$, we have
\begin{align*}
NB_F & = \sum_{\beta\in\mathbb{F}_{p^n}}\sum_{a,b\in\mathbb{F}_{p^n}}\psi_{\beta}(D_a D_b F(0))-2p^{2n}+p^n  \\
               & = p^n\order{\{(a,b)\in\mathbb{F}_{p^n}\times\mathbb{F}_{p^n}:D_a D_bF(0)=0\}}-2p^{2n}+p^n.
\end{align*}

Notice that by the definition of $NB_F$, we have
\begin{align}
NB_F & = \sum_{a \in G_1^*}\sum_{b\in G_2}\order{(D_aF)^{-1}(b)}^2-\left(\order{G_1}-1\right)\frac{\order{G_1}^2}{\order{G_2}} \notag \\
     & = \sum_{a \in G_1^*}\sum_{b\in G_2}\delta_F^2(a,b)-\left(\order{G_1}-1\right)\frac{\order{G_1}^2}{\order{G_2}} \notag \\
     & = \sum_{a\in G_1}\sum_{b\in G_2}\delta_F^2(a,b)-\order{G_1}^2-\left(\order{G_1}-1\right)\frac{\order{G_1}^2}{\order{G_2}}. \label{equ:SquareDelta}
\end{align}
Combining Propositions \ref{prop:4thfourier}, \ref{prop:autocorrelation} and \ref{prop:2ndderivative}, we have the following generalized formula, which is firstly given over finite fields with characteristic 2 by Nyberg in \cite{FSE:Nyberg94}. It provides a link between differential and linear cryptanalysis.
\begin{equation}\label{equ:DeltaFourier}
\begin{split}
\sum_{a\in G_1}\sum_{b\in G_2}\delta_F^2(a,b) & = \frac{1}{\order{G_1}\order{G_2}}\sum_{\alpha\in G_1}\sum_{\beta\in G_2}|\widehat{F}(\alpha,\beta)|^4 \\
                                              & = \frac{1}{\order{G_2}}\sum_{\alpha\in G_1}\sum_{\beta\in G_2}|\mathcal{C}_F(\alpha,\beta)|^2 \\
                                              & = \frac{1}{\order{G_2}}\sum_{\beta\in G_2}\sum_{a,b,x\in G_1}\psi_{\beta}(D_a D_b F(x)).
\end{split}
\end{equation}

For the rest of this section, we consider two special kinds of the finite Abelian groups. Firstly, we consider the case where $G_1=G_2=(\mathbb{F}_q,+)$ with odd $q$. Let $F$ be a permutation over $\mathbb{F}_q$, then we have $\mathcal{NL}(F)\geq 2(q-1)$ by Proposition \ref{prop:AmbiguityToNB}. Furthermore, if the permutation $F$ has the optimum derivative imbalance or ambiguity, it was shown \cite[Theorem 4]{TIT:PanarioSSTW13} that the nonlinearity of $F$ satisfies $\mathcal{NL}(F)\geq \frac{q-\sqrt{5q-4}}{q}$. Additionally, by Corollary \ref{cor:AmbiguityNonlinearity}, we can give an upper bound of the nonlinearity of a permutation over $\mathbb{F}_q$ with the optimum derivative imbalance or ambiguity.
\begin{cor}
  Let $G=(\mathbb{F}_q,+)$ with $q$ odd and let $F$ be a permutation over $G$ with optimum derivative imbalance or ambiguity. Then the nonlinearity of $F$ satisfies
  \[
  \frac{q-\sqrt{5q-4}}{q}\leq \mathcal{NL}(F) \leq \frac{q-\sqrt{q+4}}{q}.
  \]
\end{cor}

When the $G$ is a finite cyclic group of order $n$, we have the following similar results (the lower bound was given in \cite[Theorem 5]{TIT:PanarioSSTW13}).
\begin{cor}
  Let $G$ be a finite cyclic group of order $n$ and let $F$ be a permutation over $G$ with optimum derivative imbalance or ambiguity. Then the nonlinearity of $F$ satisfies
  \begin{enumerate}
    \item when $n$ is odd,
      \[
        \frac{n-\sqrt{5n-4}}{n}\leq \mathcal{NL}(F) \leq \frac{n-\sqrt{n+4}}{n};
      \]
    \item when $n$ is even,
      \[
        \frac{n-\sqrt{5n-6}}{n}\leq \mathcal{NL}(F) \leq \frac{n-\sqrt{n+4-\frac{4}{n-1}}}{n}.
      \]
  \end{enumerate}
\end{cor}

\section{New Results on These Indicators}\label{sec:finitefields}

In this section, we present some further results about the two indicators, $NB_F$ and $\mathcal{A}(F)$, for any map $F$ from a finite Abelian group $G_1$ to another finite Abelian group $G_2$. First, we give lower bounds of these parameters for arbitrary map $F$, as well as the functions with differential uniformity $k$. We compare our results with what are previously known and comment on one case when our bounds improves the previous results slightly. As an example, we consider the functions from $\GF{n}$ to $\GF{m}$ when $n$ is odd and $m<n$ or $n$ is even and $\frac{n}{2}<m<n$, and give a lower bound on the fourth moment of Fourier transform. We also obtain some explicit relations between deficiency and ambiguity for functions with at most $3$ values in differential spectrum. Finally, some further results are presented in the particular case of S-boxes.

\subsection{General Groups}

By Proposition \ref{prop:NBofCarlet} and Equation \eqref{eqn:nbambiguity}, we know that when $\order{G_2}$ divides $\order{G_1}$,
\begin{equation}\label{eqn:lowambiguity1}
NB_F\geq 0 \text{ or } \mathcal{A}(F)\geq (\order{G_1}-1)\order{G_1}\frac{\order{G_1}-\order{G_2}}{2\order{G_2}},
\end{equation}
with equality if and only if $F$ is perfect nonlinear. However, when $\order{G_2}$ does not divide $\order{G_1}$, there does not exist  perfect nonlinear functions between them. Next we give a general lower bound which covers the case where $\order{G_2}$ does not divide $\order{G_1}$.
\begin{prop}\label{prop:lowerbound}
  Assume that $G_1$ and $G_2$ are two finite Abelian groups. Let $F$ be a function from $G_1$ to $G_2$. Then
  \[
   NB_{F}\geq \left(\order{G_1}-1\right)\left(\left\lceil\frac{\order{G_1}^2}{\order{G_2}}\right\rceil-\frac{\order{G_1}^2}{\order{G_2}}\right),
  \]
  or equivalently,
  \[
  \mathcal{A}(F)\geq \frac{\order{G_1}-1}{2}\left(\left\lceil\frac{\order{G_1}^2}{\order{G_2}}\right\rceil-\order{G_1}\right).
  \]
\end{prop}

\begin{prf}
  Let $m=\order{G_2}$. We write $G_2=\{b_1,b_2,\dots,b_m\}$. For any fixed $a\in G_1^*$, we denote
  \[
  A_i=\{x\in G_1:D_a F(x)=b_i\},\quad 1\leq i\leq m.
  \]
  Then it is obvious that
  \[
  \sum_{i=1}^{m}\order{A_i}=\order{G_1}.
  \]
  By the Cauchy-Schwartz inequality, we have
  \begin{equation}\label{equ:numofpairs}
  \begin{split}
    \order{\{(x,y)\in G_1\times G_1:D_a F(x)=D_a F(y)\}} = &\ \order{\bigcup_{i=1}^{m}\{(x,y)\in G_1\times G_1:x,y\in A_i\}} \\
    = &\ \sum_{i=1}^{m}\order{\{(x,y)\in G_1\times G_1:x,y\in A_i\}} \\
    = &\ \sum_{i=1}^{m}\order{A_i}^2\geq\left\lceil\frac{\order{G_1}^2}{m}\right\rceil
  \end{split}
  \end{equation}
  and when $\order{G_2}$ divides $\order{G_1}$, the equality holds if and only if $\order{A_1}=\order{A_2}\cdots =\order{A_m}$, which means that the derivative $D_a F$ is balanced. Hence,
  \begin{align*}
   \sum_{a\in G_1}\order{\{(x,y)\in G_1\times G_1:D_a F(x)=D_a F(y)\}} =  & \sum_{a\in G_1^*}\order{\{(x,y)\in G_1\times G_1:D_a F(x)=D_a F(y)\}}+\order{G_1}^2 \\
    \geq & \sum_{a\in G_1^*}\left\lceil\frac{\order{G_1}^2}{m}\right\rceil+\order{G_1}^2 \\
    =    &\ (\order{G_1}-1)\left\lceil\frac{\order{G_1}^2}{m}\right\rceil+\order{G_1}^2
  \end{align*}
  and in the case where $\order{G_2}$ divides $\order{G_1}$, the equality holds if and only if all of the nonzero derivatives of $F$ are balanced, this is to say that $F$ is perfect nonlinear.

  The conclusion then follows by Corollary \ref{cor:ambiguity}. \QED
\end{prf}

When $\order{G_2}$ divides $\order{G_1}$, if a function $F$ from $G_1$ to $G_2$ achieves the lower bound of ambiguity, then its deficiency also achieves the minimum $0$. When $\order{G_2}$ does not divide $\order{G_1}$,  we obtain a nontrivial lower bound for $NB_F$ or $\mathcal{A}(F)$.

Note that if the group $G_1$ has characteristic 2, then in the case of $\order{G_1}>2\order{G_2}$, the bound of \eqref{eqn:BoundofNB} just restates the fact that $NB_F\geq 0$ since $(\order{G_1}-1)\left(2\order{G_1}-\frac{\order{G_1}^2}{\order{G_2}}\right)$ is always negative. For the particular case of vectorial Boolean functions, when $n$ is odd and $m<n$ or $n$ is even and $\frac{n}{2}<m<n$, we have $NB_F\geq 2$. Indeed, for such values of $n$ and $m$, no perfect nonlinear function exists and $NB_F$ is even. Comparing to the lower bound of \eqref{eqn:BoundofNB}, the lower bound in Proposition \ref{prop:lowerbound} does not require the knowledge of $T_F$ and characteristic of $G_1$ can be both even or odd.

Next we give another general lower bound for the indicators of differentially $k$-uniform functions, which is able to improve upon the bound in these cases. Recall that for a function $F$ from finite Abelian group $G_1$ to finite Abelian group $G_2$, by the Pigeon-Hole Principle, it is known that the differential uniformity $\Delta_F\geq \frac{\order{G_1}}{\order{G_2}}$.
\begin{prop}\label{prop:LowerBoundDiffk}
  Assume that $G_1$ and $G_2$ are two finite Abelian groups. Let $F$ be a function from $G_1$ to $G_2$ with differential uniformity $k$. Then
  \[
  NB_F\geq \left(\order{G_1}-2\right)\left\lceil\frac{\order{G_1}^2}{\order{G_2}}\right\rceil+\left\lceil\frac{(\order{G_1}-k)^2}{\order{G_2}-1}\right\rceil +k^2-\left(\order{G_1}-1\right)\frac{\order{G_1}^2}{\order{G_2}},
  \]
  or equivalently,
  \[
  \mathcal{A}(F)\geq \frac{\order{G_1}-2}{2}\left\lceil\frac{\order{G_1}^2}{\order{G_2}}\right\rceil+\frac{1}{2}\left\lceil\frac{(\order{G_1}-k)^2}{\order{G_2}-1}\right\rceil+\frac{k^2}{2}-\binom{\order{G_1}}{2}.
  \]
\end{prop}

\begin{prf}
  Let $a_0\in G_1^*$ be such that there exists $b\in G_2$ with $\delta_F(a_0,b)=k$. Without loss of generality, we may assume that $x_1,x_2,\dots,x_k$ are the $k$ different solutions in $G_1$ of equation $D_{a_0} F(x)=b$.  Denote
  \begin{align*}
  \overline{G_1}& = G_1\setminus\{x_1,x_2,\dots,x_k\},\\
  \overline{G_2}& =G_2\setminus\{b\}.
  \end{align*}
  Then by \eqref{equ:numofpairs}, we have
  \[
  \order{\{(x,y)\in\overline{G_1}\times\overline{G_1}:D_{a_0} F(x)=D_{a_0} F(y)\}}\geq \left\lceil\frac{\order{\overline{G_1}}^2}{\order{\overline{G_2}}}\right\rceil.
  \]
  Thus,
  \begin{align*}
   \order{\{(x,y)\in G_1\times G_1:D_{a_0} F(x)=D_{a_0} F(y)\}} = & \order{\{(x,y)\in G_1\times G_1:D_{a_0} F(x)=b=D_{a_0} F(y)\}} \\
          &\ +\order{\{(x,y)\in\overline{G_1}\times\overline{G_1}:D_{a_0} F(x)=D_{a_0} F(y)\}}\\
    \geq  &\ k^2+\left\lceil\frac{\order{\overline{G_1}}^2}{\order{\overline{G_2}}}\right\rceil \\
    =     &\ k^2+\left\lceil\frac{(\order{G_1}-k)^2}{\order{G_2}-1}\right\rceil.
  \end{align*}
  Finally, by \eqref{equ:numofpairs} again, we obtain
  \begin{align*}
   \sum_{a\in G_1}\order{\{(x,y)\in G_1\times G_1:D_a F(x)=D_a F(y)\}} =  &\ \sum_{a\in G_1\setminus\{0,a_0\}}\order{\{(x,y)\in G_1\times G_1:D_a F(x)=D_a F(y)\}} \\
          &\ +\order{\{(x,y)\in G_1\times G_1:D_0 F(x)=D_0 F(y)\}} \\
          &\ +\order{\{(x,y)\in G_1\times G_1:D_{a_0} F(x)=D_{a_0} F(y)\}} \\
    \geq  &\ (\order{G_1}-2)\left\lceil\frac{\order{G_1}^2}{\order{G_2}}\right\rceil+\order{G_1}^2+k^2+\left\lceil\frac{(\order{G_1}-k)^2}{\order{G_2}-1}\right\rceil.
  \end{align*}

  The lower bounds are then derived directly by Corollary \ref{cor:ambiguity}. \QED
\end{prf}

By Proposition \ref{prop:LowerBoundDiffk}, we have
\begin{align}
  NB_F & \geq \left(\order{G_1}-2\right)\left\lceil\frac{\order{G_1}^2}{\order{G_2}}\right\rceil+\left\lceil\frac{(\order{G_1}-k)^2}{\order{G_2}-1}\right\rceil +k^2-\left(\order{G_1}-1\right)\frac{\order{G_1}^2}{\order{G_2}} \notag \\
       & \geq \left(\order{G_1}-2\right)\frac{\order{G_1}^2}{\order{G_2}}+\frac{(\order{G_1}-k)^2}{\order{G_2}-1}+k^2-\left(\order{G_1}-1\right)\frac{\order{G_1}^2}{\order{G_2}} \notag \\
       & = \frac{(\order{G_1}-k)^2+(\order{G_2}-1)k^2}{\order{G_2}-1}-\frac{\order{G_1}^2}{\order{G_2}} \notag \\
       & = \frac{\order{G_2}\left(k-\frac{\order{G_1}}{\order{G_2}}\right)^2+\order{G_1}^2-\frac{\order{G_1}^2}{\order{G_2}}}{\order{G_2}-1}-\frac{\order{G_1}^2}{\order{G_2}} \notag \\
       & = \frac{\order{G_2}}{\order{G_2}-1}\left(k-\frac{\order{G_1}}{\order{G_2}}\right)^2. \label{eqn:improvedboundNB}
\end{align}
Note that this bound is always non-negative. Furthermore, when $k=\frac{\order{G_1}}{\order{G_2}}$, namely that $F$ is perfect nonlinear, it is easy to check that the equality holds.  We remark that the bound of \eqref{eqn:improvedboundNB} is better  than the bound of \eqref{eqn:BoundofNB} for some particular cases of S-boxes. For example, when $n$ is odd and $m<n$ or $n$ is even and $\frac{n}{2}<m<n$, we have $k\geq 2^{n-m}+2$. Therefore,
\[
NB_F \geq \frac{2^m}{2^m-1}(k-2^{n-m})^2\geq 4+\frac{1}{2^m-1}.
\]
Note that $NB_F$ is even, we have $NB_F \geq 6$, which slightly  improves the earlier result that $NB_F \geq 2$.

It is clear from definitions that ambiguity and deficiency are strongly correlated although they are not exactly expressed by each other in general. However, for the special case when the $\delta_F(a,b)$ of a function from $G_1$ to $G_2$ belongs to the set $\{0,i,j\}$ for any $a\in G_1^*$ and $b\in G_2$, where $1\leq i<j=\Delta_F$, there does exist an explicit relationship between them.

\begin{prop}
  Assume that $G_1$ and $G_2$ are two finite Abelian groups. Let $F$ be a function from $G_1$ to $G_2$, and $1\leq i<j=\Delta_F$. Then the following statements hold.
  \begin{enumerate}
    \item If $\mathfrak{D}_F=\{N_0,N_i\}$, then $NB_F=i(i-1)+(\order{G_1}-1)\left(\order{G_1}-\frac{\order{G_1}^2}{\order{G_2}}\right)$ or $\mathcal{A}(F)=\binom{i}{2}$, and
        \[
        \mathcal{D}(F)=\max\left\{0,(\order{G_1}-1)\left(\order{G_2}-\frac{\order{G_1}}{i}\right)\right\}.
        \]
    \item If $\mathfrak{D}_F=\{N_0,N_i,N_j\}$, then $2\mathcal{A}(F)=i\cdot j\cdot\mathcal{D}(F)+(i+j-1)\order{G_1}(\order{G_1}-1)-i\cdot j\cdot \order{G_2}(\order{G_1}-1)$.
  \end{enumerate}
\end{prop}

\begin{prf}
  By the relations \eqref{eqn:sumofN_i} and \eqref{eqn:sumofiN_i}, these results are immediate from the definitions of ambiguity and deficiency. \QED
\end{prf}

\subsection{The Particular Case of S-boxes}

In this section, we consider the functions between finite fields with characteristic 2. Some further results on these indicators are presented. Given two positive integers $n$ and $m$, when $m$ divides $n$, the \emph{trace function} from $\GF{n}$ onto its subfield $\GF{m}$ is defined as
\[
\Tr_m^n(x)=x+x^{2^m}+x^{2^{2m}}+\cdots+x^{2^{n-m}}.
\]

The following result is a refinement of Proposition \ref{prop:LowerBoundDiffk} for functions between finite fields with characteristic 2.
\begin{cor}\label{cor:DiffkUniform}
  Let $F$ be a function from $\GF{n}$ to $\GF{m}$ with differential uniformity $k$. Then
  \[
    NB_F \geq (k^2-2k)N_k+(2^n-1)(2^{n+1}-2^{2n-m}),
  \]
  or equivalently,
  \[
    \mathcal{A}(F) \geq \frac{1}{2}(k^2-2k)N_k+(2^n-1)2^{n-1}
  \]
  with equality if and only if for any $a\in\GF{n}^*$ and $b\in\GF{m}$, $\delta_F(a,b)\in\{0,2,k\}$.
\end{cor}

\begin{prf}
It is noticed that for any $a\in\GF{n}^*$ and $b\in\GF{m}$, we have $D_a F(x)=D_a F(x+a)$, therefore, $N_k=0$ for odd $k$. Then by the definition of ambiguity and equality \eqref{eqn:sumofiN_i},
\begin{align*}
     \mathcal{A}(F) & = \frac{1}{2}\sum_{i=0}^{k}i^2\times N_i-\frac{1}{2}\sum_{i=0}^{k}i\times N_i \\
                    & = \frac{1}{2}k^2\times N_k+\frac{1}{2}\sum_{i=0}^{k-2}i^2\times N_i-(2^n-1)2^{n-1} \\
                    & \geq \frac{1}{2}k^2\times N_k+\sum_{i=0}^{k-2}i\times N_i-(2^n-1)2^{n-1} \\
                    & =\frac{1}{2}k^2\times N_k+(2^n-1)2^n-k\times N_k-(2^n-1)2^{n-1} \\
                    & = \frac{1}{2}(k^2-2k)N_k+(2^n-1)2^{n-1}.
\end{align*}
The equality then comes from the simple fact that $i^2 = 2i$ if and only if $i=0$ or 2. \QED
\end{prf}

When $n=m$, the inequality \eqref{eqn:BoundofNB} gives that
\begin{equation}\label{eqn:APNBound}
NB_F \geq (2^n-1)2^{n},
\end{equation}
and this inequality is an equality if and only if $F$ is APN. This allows us to prove directly that APN functions have the optimum ambiguity.
\begin{cor}\label{cor:LowBoundofAPN}
  Let $F$ be a function over $\GF{n}$. Then
  \[
    \mathcal{A}(F) \geq (2^n-1)2^{n-1}
  \]
  and
  \[
    \mathcal{D}(F) \geq (2^n-1)2^{n-1}.
  \]
  Each of the equalities holds if and only if $F$ is APN.
\end{cor}

\begin{prf}
  For ambiguity, it is a direct consequence of \eqref{eqn:APNBound}. Now we consider the deficiency, for any $a\in\GF{n}^*$,
  \[
  \order{\{b\in\GF{n}:\delta_F(a,b)=0\}} \geq 2^{n-1}
  \]
  and the equality holds if and only if $D_a F(x)$ is 2-to-1. Thus
  \[
  \mathcal{D}(F)=\order{\{(a,b)\in\GF{n}^*\times\GF{n}:\delta_F(a,b)=0\}}=\sum_{a\in\GF{n}^*}\order{\{b\in\GF{n}:\delta_F(a,b)=0\}}\geq (2^n-1)2^{n-1}
  \]
  and the equality holds if and only if $D_a F(x)$ is 2-to-1 for any nonzero $a\in\GF{n}$. \QED
\end{prf}

For a function from $\GF{n}$ to $\GF{m}$, when $n$ is odd and $m<n$ or $n$ is even and $\frac{n}{2}<m<n$, by \eqref{bnd:ntomfunctions1}, we have
\[
 \sum_{\alpha\in\GF{n}}\sum_{\beta\in\GF{m}}|\widehat{F}(\alpha,\beta)|^4 \geq 2^{4n}+2^{3n}(2^m-1).
\]
This is the only inequality we known on the fourth moment of the Fourier transform \cite{TIT:Carlet17}. However, from the lower bound $NB_F\geq 6$ and Proposition \ref{prop:4thfourier}, we can derive an improved bound.
\begin{cor}\label{cor:LowerBoundNtoM}
  Assume that $n>2$. When $n$ is odd and $m<n$ or $n$ is even and $\frac{n}{2}<m<n$, let $F$ be a function from $\GF{n}$ to $\GF{m}$, then
  \begin{equation}\label{bud:NtoMFunction}
    \sum_{\alpha\in\GF{n}}\sum_{\beta\in\GF{m}}|\widehat{F}(\alpha,\beta)|^4 \geq 2^{4n}+2^{3n}(2^m-1)+3\cdot 2^{n+m+1}.
  \end{equation}
\end{cor}
\begin{rmk}
  In fact, the lower bound in Corollary \ref{cor:LowerBoundNtoM} is also valid for the functions from $\GF{n}$ to $\GF{n}$. Indeed, when $m=n>2$, it is easy to check that $3\cdot 2^{4n}-2^{3n+1}>2^{4n}+2^{3n}(2^n-1)+3\cdot 2^{2n+1}$. Then by the inequality \eqref{eqn:APNBound} and Proposition \ref{prop:4thfourier}, it holds that $\sum_{\alpha\in\GF{n}}\sum_{\beta\in\GF{n}}|\widehat{F}(\alpha,\beta)|^4 \geq 2^{4n}+2^{3n}(2^n-1)+3\cdot 2^{2n+1}$.
\end{rmk}

For the two special cases where $m=n-1$ or $m=n-2$, we can give a slightly improved lower bound than that in Corollary \ref{cor:LowerBoundNtoM}. Firstly, we consider the case $m=n-1$.  If $n\geq 3$ then the differential uniformity $k\geq 4$. By Corollary \ref{cor:DiffkUniform}, $NB_F \geq 8$. The following result is immediate from Proposition \ref{prop:4thfourier}.
\begin{cor}
  Assume that $n\geq 3$. Let $F$ be a function from $\GF{n}$ to $\GF{n-1}$. Then
  \[
    \sum_{\alpha\in\GF{n}}\sum_{\beta\in\GF{n-1}}\widehat{F}(\alpha,\beta)^4 \geq 3\cdot 2^{4n-1}-2^{3n}+2^{2n+2}.
  \]
\end{cor}

Similarly, in the special case $m=n-2$, we can also give the following result, which further improves upon the lower bound in Proposition \ref{prop:LowerBoundDiffk}.
\begin{prop}\label{ppn:NtoN2Function}
  Let $F$ be a function from $\GF{n}$ to $\GF{n-2}$ with differential uniformity $k$. Then
  \[
    NB_F \geq k(k-4)-2\delta_k,
  \]
  or equivalently,
  \[
    \mathcal{A}(F)\geq 3\cdot 2^{n-1}(2^n-1)+\frac{k}{2}(k-4)-\delta_k,
  \]
  where $\delta_k=0$ if $k\equiv 0\pmod 4$ and $\delta_k=2$ otherwise.
\end{prop}

\begin{prf}
  Denote $\mathcal{A}_a(F)=\order{\{(x,y)\in\GF{n}\times\GF{n}:D_a F(x)=D_aF(y) \text{ and } x\ne y\}}$. Let $a_0\in\GF{n}^*$ be such that there exists $b\in\GF{n-2}$ with $\delta_F(a_0,b)=k$. Note that it always holds that $\frac{2^n-k}{4}\leq 2^{n-2}-1$. Thus with the same notions as in Proposition \ref{prop:LowerBoundDiffk}, and similarly to the proof of Proposition \ref{prop:LowerBoundDiffk}, we have
  \begin{align*}
     \mathcal{A}_{a_0}(F) = &\ \order{\{(x,y)\in G_1\times G_1:D_{a_0} F(x)=b=D_{a_0} F(y) \text{ and } x\ne y\}}\\
                            &\  + \order{\{(x,y)\in\overline{G_1}\times\overline{G_1}:D_{a_0} F(x)=D_{a_0} F(y) \text{ and } x\ne y\}} \\
                       \geq &\  \binom{k}{2}+\frac{2^n-(k+\delta_k)}{4}\binom{4}{2}+\frac{\delta_k}{2}\binom{2}{2}=3\cdot 2^{n-1}+\frac{k}{2}(k-4)-\delta_k.
  \end{align*}
  When $a\ne 0$, $a_0$, we have $\mathcal{A}_{a}(F)\geq 2^{n-2}\binom{4}{2}$. The conclusion is immediate from the fact that $\mathcal{A}(F)=\sum_{a\in\GF{n}^*}\mathcal{A}_a(F)$. \QED
\end{prf}

Assume that $n\geq 5$, then the differential uniformity of a function from $\GF{n}$ to $\GF{n-2}$ is at least equal to $6$. By Proposition \ref{ppn:NtoN2Function}, we have $NB_F \geq 8$. By Proposition \ref{prop:4thfourier}, we have the following better bound on the fourth moment of Fourier transform for an $(n,n-2)$-function.
\begin{cor}
  Assume that $n\geq 5$. Let $F$ be a function from $\GF{n}$ to $\GF{n-2}$. Then
  \[
    \sum_{\alpha\in\GF{n}}\sum_{\beta\in\GF{n-2}}\widehat{F}(\alpha,\beta)^4 \geq 5\cdot 2^{4n-2}-2^{3n}+2^{2n+1},
  \]
\end{cor}

\begin{rmk}
  The previous two corollaries lead to the following bound for the nonlinearity of a function from $\GF{n}$ to $\GF{n-1}$ or $\GF{n-2}$,
  \begin{itemize}
    \item[1.] $n\geq 3$, for $F:\GF{n}\rightarrow\GF{n-1}$, $\mathcal{NL}(F)\leq 2^{n-1}-\frac{1}{2}\sqrt{2^n+\frac{4}{2^{n-1}-1}}$,
    \item[2.] $n\geq 5$, for $F:\GF{n}\rightarrow\GF{n-2}$, $\mathcal{NL}(F)\leq 2^{n-1}-\frac{1}{2}\sqrt{2^n+\frac{1}{2^{n-2}-1}}$,
  \end{itemize}
  which means the well-known fact that the covering radius bound is not tight. Some recent results on the nonlinearity of APN functions and some characterizations of the differential uniformity of vectorial functions by the Walsh transform are given in \cite{TIT:Carlet17}, and results on the covering radius bound for those $(n,m)$-functions that are sufficiently unbalanced or satisfy some conditions are given in \cite{TIT:XuCMW18}.
\end{rmk}

For the rest of this  section, we consider two kinds of special functions, power functions and plateaued functions, for their important applications in sequence and cryptography. The \emph{plateaued functions} are those Boolean functions whose squared Fourier transform takes one single nonzero value. \emph{Vectorial plateaued Functions} are functions whose component functions are plateaued.

For the inverse function $F(x)=x^{-1}$ over $\GF{n}$, when $n$ is odd, it is well known that the function is an APN permutation, and it has the optimum derivative imbalance or ambiguity. When $n$ is even, it was proved in \cite{FFA:CarletD07} that $NB_F=(2^n-1)(2^n+8)$  (note there is a typo in Example 1  \cite{FFA:CarletD07}). Indeed, the inverse function has the lowest derivative imbalance or ambiguity among all the power permutations (see Remark \ref{rmk:InverseFunction} below).

Let $F(x)=x^d$ be a power function from $\GF{n}$ to $\GF{m}$, we can deduce that
\[
\begin{split}
  \sum_{a\in\GF{n}^*}\sum_{b\in\GF{m}}\delta_F^2(a,b)=\sum_{a\in\GF{n}^*}\sum_{b\in\GF{m}}\delta_F^2\left(1,\frac{b}{a^d}\right)=\sum_{a\in\GF{n}^*}\sum_{b\in\GF{m}}\delta_F^2(1,b) =(2^n-1)\sum_{b\in\GF{m}}\delta_F^2(1,b).
\end{split}
\]
Then by \eqref{equ:SquareDelta}, it is easy to see the following corollary.
\begin{cor}
  Let $F(x)=x^d$ be a power function from $\GF{n}$ to $\GF{m}$. Then
  \[
   NB_F=(2^n-1)\sum_{b\in\GF{m}}|(D_1 F)^{-1}(b)|^2-2^{2n-m}(2^n-1),
  \]
  or equivalently,
  \[
   \mathcal{A}(F)=\frac{2^n-1}{2}\sum_{b\in\GF{m}}|(D_1 F)^{-1}(b)|^2-(2^n-1)2^{n-1}.
  \]
\end{cor}

An immediate result from Corollary \ref{cor:DiffkUniform} is given.
\begin{cor}
  Let $F(x)=x^d$ be a power function from $\GF{n}$ to $\GF{m}$ with differential uniformity $k$. Then
  \[
    NB_F \geq (k^2-2k)(2^n-1)N_k'+(2^n-1)(2^{n+1}-2^{2n-m}),
  \]
  or equivalently,
  \[
    \mathcal{A}(F) \geq \frac{1}{2}(k^2-2k)(2^n-1)N_k'+(2^n-1)2^{n-1}
  \]
  where $N_k'=\order{\{b\in\GF{m}:\delta_F(1,b)=k\}}$. And the equality holds if and only if for any $b\in\GF{m}$, $\delta_F(1,b)\in\{0,2,k\}$.
\end{cor}

\begin{rmk}\label{rmk:InverseFunction}
  It is well known that for any power permutation $F$ over $\GF{n}$ with even $n$, the differential uniformity is at least 4 \cite{DAM:Hou06}. Thus we have $NB_F \geq (4^2-2\times 4)(2^n-1)+(2^n-1)(2^{n+1}-2^{2n-n})=(2^n-1)(2^n+8)$.
\end{rmk}

For plateaued functions, we have the following results.

\begin{cor}
  Let $F$ be an $n$ variables Boolean function, which is plateaued of amplitude $\mu$. Then
  \[
    NB_F = 2^{n-1}(\mu^2-2^n),
  \]
  or equivalently,
  \[
  \mathcal{A}(F)=2^{n-2}(\mu^2-2^n)+2^{n-2}(2^n-1)(2^n-2).
  \]
\end{cor}

\begin{cor}
  Let $F$ be a function from $\GF{n}$ to $\GF{m}$ such that all component functions $f_{\beta}:x\in\GF{n}\mapsto\Tr_1^m(\beta F(x))$, $\beta\in\GF{m}^*$ are plateaued of amplitude $\mu_{\beta}$. Then
  \[
    NB_F = 2^{n-m}\sum_{\beta\in\GF{m}^*}\mu_{\beta}^2-2^{2n-m}(2^m-1),
  \]
  or equivalently,
  \[
  \mathcal{A}(F)=2^{n-m-1}\sum_{\beta\in\GF{m}^*}\mu_{\beta}^2+2^{3n-m-1}-2^{n-1}(2^{n+1}-1).
  \]
\end{cor}

For a power function $F(x)=x^d$, Carlet proved the following results which can make the characterization of these two parameters easier.

\begin{lem}[See \cite{TIT:Carlet15}]\label{lem:PowerPlateaued}
  Let $F(x) = x^d$ be any power function over $\GF{n}$. Then for every $v\in\GF{n}$, every $x\in\GF{n}$, and every $\lambda\in\GF{n}^*$ we have
  \begin{align*}
    \order{\{(a,b)\in\GF{n}\times\GF{n}:D_a F(b)+D_a F(x)=v\}}=\order{\{(a,b)\in\GF{n}\times\GF{n}:D_a F(b)+D_a F\left(\frac{x}{\lambda}\right)=\frac{v}{\lambda^d}\}}.
  \end{align*}
  Moreover, $F$ is plateaued if and only if, for every $v\in\GF{n}$:
  \begin{align*}
    \order{\{(a,b)\in\GF{n}\times\GF{n}:D_a F(b)+D_a F(1)=v\}}=\order{\{(a,b)\in\GF{n}\times\GF{n}:D_a F(b)+D_a F(0)=v\}}.
  \end{align*}
\end{lem}

Note that the previous lemma is proved for $(n,n)$-functions which are power and plateaued, we remark that the result is also valid for $(n,m)$-functions which are power and plateaued.

\begin{cor}
  Let $F(x) = x^d$ be any power function from $\GF{n}$ to $\GF{m}$. If $F$ is also plateaued, then
  \[
    NB_F = 2^{n}(2^n-1)\left(\order{(D_1F)^{-1}(1)}-1\right)-2^{2n-m}(2^n-1)=2^{n}(2^n-1)(\delta_F(1,1)-1)-2^{2n-m}(2^n-1),
  \]
  or equivalently,
  \[
  \mathcal{A}(F)=2^{n-1}(2^n-1)\left(\order{(D_1F)^{-1}(1)}-1\right)=2^{n-1}(2^n-1)(\delta_F(1,1)-1).
  \]
\end{cor}

\begin{prf}
  By Corollary \ref{cor:ambiguity} and Lemma \ref{lem:PowerPlateaued}, we have
  \begin{align*}
    NB_F           = &\ \sum_{a\in\GF{n}}\order{\{(b,x)\in\GF{n}\times\GF{n}:D_a F(b)=D_a F(x)\}}-2^{2n}-2^{2n-m}(2^{n}-1) \\
                   = &\ \sum_{x\in\GF{n}}\order{\{(a,b)\in\GF{n}\times\GF{n}:D_a F(b)=D_a F(x)\}}-2^{2n}-2^{2n-m}(2^{n}-1) \\
                   = &\ \sum_{x\in\GF{n}^*}\order{\{(a,b)\in\GF{n}\times\GF{n}:D_a F(b)=D_a F(x)\}} \\
                     &\ +\order{\{(a,b)\in\GF{n}\times\GF{n}:D_a F(b)=D_a F(0)\}}-2^{2n}-2^{2n-m}(2^{n}-1) \\
                   = &\ \sum_{x\in\GF{n}^*}\order{\{(a,b)\in\GF{n}\times\GF{n}:D_a F(b)=D_a F(1)\}} \\
                     &\ +\order{\{(a,b)\in\GF{n}\times\GF{n}:D_a F(b)=D_a F(0)\}}-2^{2n}-2^{2n-m}(2^{n}-1) \\
                   = &\ 2^{n}\order{\{(a,b)\in\GF{n}\times\GF{n}:D_a F(b)=D_a F(0)\}}-2^{2n}-2^{2n-m}(2^{n}-1) \\
                   = &\ 2^{n}\order{\{(a,b)\in\GF{n}^*\times\GF{n}:D_a F(b)=D_a F(0)\}}-2^{2n-m}(2^n-1) \\
                   = &\ 2^{n}\order{\{(a,b)\in\GF{n}^*\times\GF{n}:D_1 F(b)=D_1 F(0)\}}-2^{2n-m}(2^n-1) \\
                   = &\ 2^{n}(2^n-1)\order{\{b\in\GF{n}:D_1 F(b)=1\}}-2^{2n-m}(2^n-1).
  \end{align*}
  The proof is completed. \QED
\end{prf}

\begin{example}\label{exa:PowerPlateaued}
  When $F$ is a quadratic power function $x\mapsto x^{2^i+2^j}$ over $\GF{n}$ where $i>j$. Then $D_1F(x)=(x+1)^{2^i+2^j}+x^{2^i+2^j}=x^{2^i}+x^{2^j}+1=1$ if and only if $x^{2^{i-j}}=x$, which is equivalent to $x\in\GF{s}$, where $s=\gcd(i-j,n)$. We have $NB_F=2^{n}(2^n-1)(2^s-1)$. The deficiency can be easily to obtain, $\mathcal{D}(F)=(2^n-1)(2^{n}-2^{n-s})$.
\end{example}

\section{Conclusions and Some Open Problems}\label{sec:conclusion}

In this paper, we studied the non-balancedness of the derivatives of functions between any two finite Abelian groups with possible different orders. We systematically compared two parameters appeared in the literature and observed that the parameter called ambiguity is equivalent to an indicator (that we called derivative imbalance) introduced earlier by Carlet and Ding in the study of the nonlinearity of S-boxes. We gave lower bounds on these parameters for these general maps, and in the particular case of differentially $k$-uniform functions. We generalized a characterization of these parameters by the fourth moment of Fourier transform. We also investigated the connections between these parameters and the behavior of derived functions such as second-order derivatives and autocorrelation functions. Moreover, when the groups are the finite fields with characteristic 2, some further results were presented.

We gave some new lower bounds on the fourth moment of Fourier transform by analyzing the lower bounds of the ambiguity of a function from $\GF{n}$ to $\GF{m}$ when $n$ is odd and $m<n$ or $n$ is even and $\frac{n}{2}<m<n$. As consequences. we obtained $NB_F \geq 6$ or $8$ in some cases. This has improved the previous lower bound $NB_F \geq 2$. However, in order to obtain an upper bound  for nonlinearity that is better than the covering radius bound, we need to show that $NB_F > 2^{n-m+2}(2^{n/2}+1)(2^m-1)$. This is a well-known open problem and worthy of further study \cite{FFA:CarletD07,WAIFI:Carlet14}.

\section*{Acknowledgements}

We want to thank Cunsheng Ding, Sihem Mesnager, and anonymous reviewers for helpful suggestions. This work was supported by the National Natural Science Foundation of China (Grant No. 61572491 and 11688101) and Science and Technology on Communication Security Laboratory (Grant No. 6142103010701). Research of Qiang Wang was partially supported by NSERC of Canada.


{\footnotesize
\newcommand{\etalchar}[1]{$^{#1}$}

}

\end{document}